# Quantum Heat Engines and the Generalized Uncertainty Principle

**Gardo Blado, Jonathan Nguyen, Giovani Renteria, Skylar Gay, Bryce Mortimer**

*Physics Discipline*
*College of Science and Engineering*
*Houston Baptist University*
*7502 Fondren Rd., Houston, Texas, U.S.A*

## Abstract

We study the effects of the generalized uncertainty principle (GUP) on the efficiency of quantum heat engines based on a particle in an infinite square well using the partition function approach. In particular, we study the Carnot and Otto heat engines. For the system we used, the GUP-corrected efficiencies turned out to be lower than efficiencies without the GUP effects. However, as expected, GUP effects increase as the temperature of the cold heat bath decreases and as the width of the potential well decreases.

# 1. Introduction

In the past decade a number of papers have been published that makes quantum gravity effects via the generalized uncertainty principle (GUP) accessible to undergraduate students in physics [1, 2, 3]. These papers discussed topics covered in the undergraduate physics curriculum such as quantum wells, WKB approximation applied to various phenomena and a two-state system. In addition, there is a very rich literature which discuss the GUP applied to various systems commonly discussed in a standard quantum mechanics class such as the simple harmonic oscillator [4], the delta function [5], and the hydrogen atom [6, 7] among others. The present paper will explore the application of the GUP to quantum thermodynamics by studying GUP-effects on the efficiency of the most commonly discussed heat engines in undergraduate physics namely the Carnot and Otto engines. The combination of concepts in quantum mechanics, thermodynamics and quantum gravity results in this paper can be of pedagogical value for undergraduate students.

The paper is organized as follows. A general discussion on how to calculate the efficiency of a quantum heat engine using the partition function of a particle in an infinite square well is done in section 2. The general formula derived in section 2 to calculate the heat involved in the processes is then applied to the quantum Carnot and quantum Otto cycles in sections 3 and 4 respectively to derive known results. Section 5 derives a corresponding general formula to compute the GUP-corrected heat energy. This general formula is applied to get the GUP-corrected efficiencies of the quantum Carnot and Otto cycles in sections 6 and 7 respectively. Finally, some conclusions are discussed in the last section.

## 2. Quantum Heat Engine

Let us first outline the approach we will take in calculating the efficiency. Our approach follows [8] and [9]. We will calculate the efficiency as.

*Equation 1:* $\eta = \frac{W}{Q_{in}} = \frac{Q_{in} - |Q_{out}|}{Q_{in}} = 1 - \frac{|Q_{out}|}{Q_{in}}$.

The heat involved is calculated using the entropy,

*Equation 2:* $\Delta S = \int \frac{dQ}{T}$.

To calculate the entropy, we use the partition function $Z$.

*Equation 3:* $Z \equiv \sum_{n=1}^{\infty} e^{-\beta E_n}$

with the $\beta$ parameter defined by

*Equation 4:* $\beta \equiv \frac{1}{kT}$

where $k$ is the Boltzmann's constant. From $Z$, we calculate the free energy

*Equation 5:* $F = -\frac{1}{\beta} \ln Z$

and internal energy

*Equation 6:* $U = -\frac{\partial}{\partial \beta} \ln Z$.

From $F$ and $U$, we get the entropy [10, 9]

*Equation 7:* $S = k\beta(U - F)$.

Let us now consider a quantum heat engine based on a particle confined in an infinite square well (ISW) potential. The energy of a particle in an ISW [11] is

*Equation 8:* $E_n = \gamma n^2$

where we define

*Equation 9:* $\gamma \equiv \frac{\pi^2 \hbar^2}{2mL^2}$

with $m$ the mass of the particle in the well of width $L$ and $n = 1,2,3, \ldots$. From Equation 3, for the infinite square well, we have $Z = \sum_{n=1}^{\infty} e^{-\beta \gamma n^2}$. We make the approximation [8],

*Equation 10:* $= \sum_{n=1}^{\infty} e^{-\beta \gamma n^2} \approx \int_0^{\infty} dn\, e^{-\beta \gamma n^2} = \frac{1}{2}\sqrt{\frac{\pi}{\beta \gamma}}$.

Using this partition function, we get from Equation 3, Equation 5, Equation 6, and Equation 7, the entropy

*Equation 11:* $S = \frac{k}{2} + k \ln\left(\frac{1}{2}\sqrt{\frac{\pi}{\beta \gamma}}\right) = S_0 - \frac{1}{2} k \ln(\beta \gamma) = \frac{k}{2} + k \ln\left(\frac{1}{2}\sqrt{\frac{\pi}{\beta \gamma}}\right)$

with the constant, $S_0 \equiv \frac{k}{2} + k \ln\left(\frac{\sqrt{\pi}}{2}\right)$. From Equation 2 and Equation 11, let us calculate the heat involved when a particle in an infinite square well goes from an initial state $i$ to a final state $f$. From Equation 2, we have

*Equation 12:* $dQ = \frac{dS}{k\beta}$

where the entropy is a function of $\beta$ and $\gamma$, $S(\beta,\gamma)$. Hence $dS = \frac{\partial S}{\partial \beta}d\beta + \frac{\partial S}{\partial \gamma}d\gamma$ and from Equation 11, we get $dS = -\frac{k}{2\beta}d\beta - \frac{k}{2\gamma}d\gamma$. Hence Equation 12 becomes $dQ = -\frac{d\beta}{2\beta^2} - \frac{d\gamma}{2\beta\gamma}$. Integrating the preceding equation from an initial state $i$ to a final state $f$, we get

Equation 13: $Q_{if} = \frac{1}{2}\left(\frac{1}{\beta_f} - \frac{1}{\beta_i}\right) - \frac{1}{2}\int_i^f \frac{d\gamma}{\beta\gamma}$.

Let us next use Equation 13 to find the efficiencies of the quantum Carnot and quantum Otto engines involving a particle in an infinite square well. For future reference, we note that we impose a condition [9, 8]

Equation 14: $\beta\gamma = k_0 =$ constant

in the adiabatic process, which is apparently consistent with Equation 11 ($\Delta S = 0$).

## 3. Quantum Carnot Cycle

The classical Carnot cycle is illustrated in Figure 1. Processes AB and CD are isothermal processes at temperatures $T_h$ and $T_l$ respectively with $T_h > T_l$. The heat $Q_h$ enters the system at the process AB while the heat loss of $Q_l$ occurs in the process CD. Processes BC and DA are adiabatic expansion and compression processes respectively. In introductory physics classes the Carnot cycle is discussed as the most efficient heat engine with an efficiency of $\eta_{classical\ Carnot} = 1 - \frac{T_l}{T_h}$. In transitioning to the quantum Carnot cycle, we use the equivalent isothermal and adiabatic processes as in the classical Carnot cycle [8].

For the Carnot cycle, Equation 1 becomes $\eta = 1 - \frac{|Q_{out}|}{Q_{in}} = 1 - \frac{|Q_l|}{Q_h}$. Let us calculate $Q_l$ and $Q_h$ for the Carnot cycle. For an isothermal process, we get from Equation 13.

Equation 15: $Q_{isothermal} = -\frac{1}{2\beta}\int_i^f \frac{d\gamma}{\gamma} = -\frac{1}{2\beta}\ln\left(\frac{\gamma_f}{\gamma_i}\right)$.

Hence for AB and CD:

Equation 16: $Q_{AB} = -\frac{1}{2\beta_h}\ln\left(\frac{\gamma_B}{\gamma_A}\right) = -\frac{1}{\beta_h}\ln\left(\frac{L_A}{L_B}\right) = \frac{1}{\beta_h}\ln\left(\frac{L_B}{L_A}\right)$

and

Equation 17: $Q_{CD} = -\frac{1}{2\beta_l}\ln\left(\frac{\gamma_D}{\gamma_C}\right) = -\frac{1}{\beta_l}\ln\left(\frac{L_C}{L_D}\right)$

where we used Equation 9. For the adiabatic process $Q_{BC}$, Equation 13 gives

*Equation 18:* $Q_{BC} = \frac{1}{2}\left(\frac{1}{\beta_l} - \frac{1}{\beta_h}\right) - \frac{1}{2}\int_B^C \frac{d\gamma}{\beta\gamma}$.

Evaluating the integral using Equation 14, $\frac{1}{2}\int_B^C \frac{d\gamma}{\beta\gamma} = \frac{1}{2k_0}(\gamma_C - \gamma_B) = \frac{\gamma_C}{2k_0} - \frac{\gamma_B}{2k_0}$. But $k_0 = \beta_C\gamma_C = \beta_B\gamma_B$ so we can write $\frac{1}{2}\int_B^C \frac{d\gamma}{\beta\gamma} = \frac{\gamma_C}{2\beta_C\gamma_C} - \frac{\gamma_B}{2\beta_B\gamma_B} = \frac{1}{2}\left(\frac{1}{\beta_C} - \frac{1}{\beta_B}\right) = \frac{1}{2}\left(\frac{1}{\beta_l} - \frac{1}{\beta_h}\right)$. Hence, Equation 18 yields

*Equation 19:* $Q_{BC} = 0$.

Similarly, we can show

*Equation 20:* $Q_{DA} = 0$.

Note that from Equation 16 and Equation 17 $Q_{AB} = Q_h > 0$ since there is an expansion from $A \to B$ while $Q_{CD} = Q_l < 0$ since there is a compression from $C \to D$. Hence from Equation 1, Equation 16 and Equation 17 we get

*Equation 21:* $\eta = 1 - \frac{|Q_l|}{Q_h} = 1 - \frac{|Q_{CD}|}{Q_{AB}} = 1 - \frac{\frac{1}{\beta_l}\ln\left(\frac{L_C}{L_D}\right)}{\frac{1}{\beta_h}\ln\left(\frac{L_B}{L_A}\right)} = 1 - \frac{T_l}{T_h}\frac{\ln\left(\frac{L_C}{L_D}\right)}{\ln\left(\frac{L_B}{L_A}\right)}$.

Note for the adiabatic processes $\beta_C\gamma_C = \beta_B\gamma_B$ and $\beta_D\gamma_D = \beta_A\gamma_A$ and for the isothermal processes, $\beta_A = \beta_B$ and $\beta_C = \beta_D$. Hence, $\frac{\gamma_D}{\gamma_C} = \frac{\gamma_A}{\gamma_B}$. With Equation 9, we get $\frac{L_C}{L_D} = \frac{L_B}{L_A}$. This yields from Equation 21, the quantum Carnot efficiency

*Equation 22:* $\eta_C = 1 - \frac{T_l}{T_h} = 1 - \frac{\beta_h}{\beta_l}$

which is also the classical Carnot cycle efficiency [8].

## 4. Quantum Otto Cycle

The classical Otto cycle is illustrated in Figure 2 with $T_B > T_A > T_C > T_D$. Processes AB and CD are isochoric (constant volume) processes at volumes $V_h$ and $V_l$ respectively with $V_l > V_h$. The heat $Q_h$ enters the system at the process AB while the heat loss of $Q_l$ occurs in the process CD. Processes BC and DA are adiabatic expansion and compression processes respectively. The efficiency of a classical Otto engine is $\eta_{classical\ Otto} = 1 - \left(\frac{V_h}{V_l}\right)^{\gamma_A - 1}$ with $\gamma_A$ is the adiabatic exponent of the ideal gas working substance. In transitioning to the quantum Otto cycle, we use the equivalent isothermal and adiabatic processes as in the classical Carnot cycle [8].

From Equation 13, let us calculate the efficiency of a quantum Otto cycle involving a particle in an infinite square well. For the isochoric process as in Figure 2, $A \to B$ (constant length of the ISW), we have (with $\beta_B = \beta_h, \beta_D = \beta_l, \gamma_B = \gamma_h$ and $\gamma_D = \gamma_l$ as in [8]),

Equation 23: $Q_{AB} = \frac{1}{2}\left(\frac{1}{\beta_B} - \frac{1}{\beta_A}\right) = \frac{1}{2}\left(\frac{1}{\beta_h} - \frac{1}{\beta_A}\right).$

With Equation 14,

Equation 24: $\beta_A \gamma_A = \beta_D \gamma_D = \beta_l \gamma_l.$

Hence $\beta_A = \frac{\beta_l \gamma_l}{\gamma_A} = \frac{\beta_l \gamma_l}{\gamma_h}$. Equation 23 becomes $Q_{AB} = \frac{1}{2}\left(\frac{1}{\beta_h} - \frac{\gamma_h}{\beta_l \gamma_l}\right)$ or

Equation 25: $Q_{AB} = \frac{\gamma_h}{2}\left(\frac{1}{\beta_h \gamma_h} - \frac{1}{\beta_l \gamma_l}\right).$

Similarly, we can show that for the other isochoric process,

Equation 26: $Q_{CD} = \frac{\gamma_l}{2}\left(\frac{1}{\beta_l \gamma_l} - \frac{1}{\beta_h \gamma_h}\right).$

For the adiabatic process $B \to C$, from Equation 13 and Equation 14, we get $Q_{BC} = \frac{1}{2}\left(\frac{1}{\beta_C} - \frac{1}{\beta_B}\right) - \frac{1}{2k_0}\int_B^C d\gamma = \frac{1}{2}\left(\frac{1}{\beta_C} - \frac{1}{\beta_B}\right) - \frac{1}{2k_0}(\gamma_C - \gamma_B)$. From Equation 14 we can write $\frac{1}{2k_0}(\gamma_C - \gamma_B) = \frac{\gamma_C}{2k_0} - \frac{\gamma_B}{2k_0}.$

But for the process $B \to C$,

Equation 27: $k_0 = \beta_C \gamma_C = \beta_B \gamma_B = \beta_h \gamma_h.$

So, $\frac{1}{2k_0}(\gamma_C - \gamma_B) = \frac{\gamma_C}{2\beta_C \gamma_C} - \frac{\gamma_B}{2\beta_B \gamma_B} = \frac{1}{2}\left(\frac{1}{\beta_C} - \frac{1}{\beta_B}\right)$. Hence,

Equation 28: $Q_{BC} = 0.$

Similarly, we can show that for the other adiabatic process $D \to A$, with $\beta_A \gamma_A = \beta_D \gamma_D$

Equation 29: $Q_{DA} = 0.$

Equation 25 to Equation 29, agree with [8] with $Q_{AB} = Q_{in}$ and $Q_{CD} = Q_{out}$. Hence from Equation 1, Equation 25, Equation 26, we get $\eta_O = 1 - \frac{|Q_{out}|}{Q_{in}} = 1 - \frac{|Q_{CD}|}{Q_{AB}} = 1 - \frac{\gamma_l}{\gamma_h}$ and with Equation 9, we get the quantum Otto efficiency

Equation 30: $\eta_O = 1 - \frac{\gamma_l}{\gamma_h} = 1 - \frac{L_h^2}{L_l^2}$

as also derived in [8]. We next turn to the calculation of the GUP-corrected efficiencies of the quantum Carnot and Otto cycles using a similar method.

## 5. GUP-corrected Quantum Heat Engine

The generalized uncertainty principles has been discussed widely in the literature [12, 13, 14, 15, 16, 17, 18, 19]. Quantum gravity theories lead to a minimal length scale which can result from a modification of the Heisenberg uncertainty principle $\Delta x \Delta p \geq \frac{\hbar}{2}$ to a generalized uncertainty principle (GUP)

Equation 31: $\Delta x \Delta p \geq \frac{\hbar}{2} F(\beta_G, \Delta p)$

[12, 20, 21, 22, 23] where $F(\beta_G, \Delta p)$ is some function of the the GUP parameter $\beta_G$ and the uncertainty in momentum $\Delta p$. In [1], we employ a first order approximation of Equation 31

Equation 32: $\Delta x \Delta p \geq \frac{\hbar}{2}(1 + \beta_G (\Delta p)^2)$.

A modified momentum-position commutation relations given by $[x, p] = i\hbar(1 + \beta_G p^2)$ replaces $[x_0, p_0] = i\hbar$. The modified commutation relation can be shown to be satisfied by $x = x_0$ and $p = p_0(1 + \beta_G p_0^2)$ with $p_0 = \frac{\hbar}{i}\frac{d}{dx}$. The new operators lead to a GUP-corrected Schrodinger equation, $\left(\frac{p_0^2}{2m} + \frac{\beta_G}{m} p_0^4 + V\right)\psi = E\psi$. From this modified Schrodinger equation the energy eigenvalues of a particle in an infinite square well can be derived and is given by

Equation 33: $E_n^G = \frac{n^2 \pi^2 \hbar^2}{2mL^2} + \frac{n^4}{L^4}\frac{\beta_G \pi^4 \hbar^4}{m}$

for a potential well of width $L$ [1, 13].

In this section, we derive the GUP-corrected $Q_{if}^G$ version of Equation 13. Similar to Equation 3, we can write the GUP-corrected partition function as $Z^G \equiv \sum_{n=1}^{\infty} e^{-\beta E_n^G}$. From Equation 33 the GUP-corrected energy is given by $E_n^G = E_n + \frac{n^4}{L^4}\frac{\beta_G \pi^4 \hbar^4}{m} = E_n(1 + \delta n^2)$ with (using also Equation 9)

Equation 34: $\delta = \delta(\beta_G, L) = \frac{2\beta_G \pi^2 \hbar^2}{L^2} = 4m\beta_G \gamma \ll 1$

with a small GUP parameter $\beta_G$. We have $Z^G \equiv \sum_{n=1}^{\infty} e^{-\beta E_n} e^{-\beta E_n \delta n^2} \approx \sum_{n=1}^{\infty} e^{-\beta E_n}(1 - \delta \beta E_n n^2) = \sum_{n=1}^{\infty} e^{-\beta E_n} - \delta \beta \sum_{n=1}^{\infty} E_n n^2 e^{-\beta E_n}$. $Z^G \approx Z - \delta \beta \sum_{n=1}^{\infty} E_n n^2 e^{-\beta E_n}$. From Equation 8 and Equation 9, we have $Z^G \approx Z - \delta \beta \sum_{n=1}^{\infty} \gamma n^2 n^2 e^{-\beta E_n} = Z - \delta \beta \gamma \underbrace{\sum_{n=1}^{\infty} n^4 e^{-\beta \gamma n^2}}_{\approx \int_0^{\infty} dn\, n^4 e^{-\beta \gamma n^2}} \approx Z - \delta \beta \gamma \left(\frac{3\sqrt{\pi}}{8}(\beta \gamma)^{-5/2}\right)$. We used the approximation in Equation 10. With Equation 34, we get

$Z^G \approx Z - 4m\beta_G \gamma \beta \gamma \left(\frac{3\sqrt{\pi}}{8}(\beta \gamma)^{-5/2}\right)$ or

Equation 35: $Z^G \approx Z - K(\beta^3 \gamma)^{-1/2}$

where

*Equation 36:* $K \equiv \frac{3\sqrt{\pi}}{2}\beta_G m$.

Since $K \sim \beta_G$ and is hence small, we can make the approximation (using also Equation 10),

*Equation 37:* $\ln Z^G = \ln\left(Z\left[1 - \frac{K(\beta^3\gamma)^{-1/2}}{Z}\right]\right) \approx \ln Z - \frac{2K}{\sqrt{\pi}\beta}$.

Similar to the calculations in Section 2 above, we then calculate the GUP-corrected free energy $F^G$, internal energy $U^G$, entropy $S^G$ and finally the heat $Q_{if}^G$. We just list them next. Note that the quantities without the superscript "G" are the values without the GUP correction.

$F^G = -\frac{1}{\beta}\ln Z^G = F + \frac{2K}{\sqrt{\pi}\beta^2}$

$U^G = -\frac{\partial}{\partial \beta}\ln Z^G = U - \frac{2K}{\sqrt{\pi}\beta^2}$

$S^G = k\beta(U^G - F^G) = S - \frac{4kK}{\sqrt{\pi}\beta}$

What is notable about Equation 37 is that the GUP correction terms are independent of the width of the square well, in agreement to the analogous independence from the volume of an ideal gas in the GUP-corrected thermodynamic quantities in [24]. Similar to Equation 12, we have $dQ^G = \frac{dS^G}{k\beta} = \frac{dS}{k\beta} - \frac{1}{k\beta}d\left(\frac{4kK}{\sqrt{\pi}\beta}\right) = \frac{dS}{k\beta} - \frac{1}{k\beta}d\left(\frac{6\beta_G mk}{\beta}\right)$ so integrating from an initial state $i$ to a final state $f$, we get $Q_{if}^G = Q_{if} - 6\beta_G m \int_i^f \frac{d(\beta^{-1})}{\beta}$. This yields

*Equation 38:* $Q_{if}^G = Q_{if} - \frac{\lambda}{2}\left(\frac{1}{\beta_f^2} - \frac{1}{\beta_i^2}\right)$, $\lambda \equiv 6\beta_G m$.

With $Q_{if}$ given in Equation 13.

## 6. GUP-corrected Quantum Carnot Cycle

Recall the processes in a Carnot engine: $A \to B$, an isothermal expansion at temperature $T_h$, $B \to C$, an adiabatic expansion, $C \to D$, an isothermal compression at a lower temperature $T_l$, and an adiabatic compression, $D \to A$. From Equation 38, for the isothermal processes (equal initial and final temperatures), $Q_{AB}^G = Q_{AB} > 0$ and $Q_{CD}^G = Q_{CD} < 0$. For the corresponding "adiabatic" processes, $Q_{BC}^G = -\frac{\lambda}{2}\left(\frac{1}{\beta_C^2} - \frac{1}{\beta_B^2}\right) = \frac{\lambda}{2}\left(\frac{1}{\beta_B^2} - \frac{1}{\beta_C^2}\right) > 0$ since $T_B = T_h > T_C = T_l$. Similarly $Q_{DA}^G = -\frac{\lambda}{2}\left(\frac{1}{\beta_A^2} - \frac{1}{\beta_D^2}\right) < 0$ since $T_A = T_h > T_D = T_l$. With the GUP correction the processes $B \to C$ and $D \to A$ have some heat involved. Hence,

*Equation 39:* $Q_{in}^G = Q_{AB}^G + Q_{BC}^G = Q_{AB} + \frac{\lambda}{2}\left(\frac{1}{\beta_h^2} - \frac{1}{\beta_l^2}\right)$

and

*Equation 40:* $Q_{out}^G = Q_{CD}^G + Q_{DA}^G = Q_{CD} - \frac{\lambda}{2}\left(\frac{1}{\beta_h^2} - \frac{1}{\beta_l^2}\right) = -|Q_{CD}| - \frac{\lambda}{2}\left(\frac{1}{\beta_h^2} - \frac{1}{\beta_l^2}\right).$

From Equation 1,

*Equation 41:* $\eta_C^G = 1 - \frac{|Q_{out}^G|}{Q_{in}^G} = 1 - \frac{|Q_{CD}| + \Delta Q}{Q_{AB} + \Delta Q},$

where we define

*Equation 42:* $\Delta Q \equiv \frac{\lambda}{2}\left(\frac{1}{\beta_h^2} - \frac{1}{\beta_l^2}\right)$

Note that $\Delta Q \ll Q_{if}$ in Equation 38 since $\lambda \sim \beta_G$. In addition, $\Delta Q > 0$ due to Equation 4. Let us manipulate the second term in Equation 41 which has the relatively small quantity $\Delta Q$. $\frac{|Q_{CD}| + \Delta Q}{Q_{AB} + \Delta Q} = \frac{|Q_{CD}|\left(1 + \frac{\Delta Q}{|Q_{CD}|}\right)}{Q_{AB}\left(1 + \frac{\Delta Q}{Q_{AB}}\right)} \approx \frac{|Q_{CD}|}{Q_{AB}}\left(1 + \frac{\Delta Q}{|Q_{CD}|}\right)\left(1 - \frac{\Delta Q}{Q_{AB}}\right)$. Expanding this expression to the first order of $\Delta Q$ (with $\Delta Q \sim \lambda \sim \beta_G$, Equation 41 becomes $\eta_C^G = 1 - \frac{|Q_{CD}|}{Q_{AB}} - \Delta Q \left(\frac{Q_{AB} - |Q_{CD}|}{Q_{AB}^2}\right)$. With Equation 21, we finally get

*Equation 43:* $\eta_C^G = \eta_C - \Delta\eta,\ \Delta\eta \equiv \frac{\Delta Q}{Q_{AB}}\eta_C.$

Note that the GUP correction term $\Delta\eta > 0$ since $\Delta Q$ and $Q_{AB}$ are both positive as pointed out above. Hence the GUP-corrected quantum Carnot efficiency is less than the non-GUP quantum Carnot efficiency. Let us rewrite Equation 43 in terms of the ratios of the higher and lower temperatures and highest and lowest lengths. Remember that $Q_{AB}$ is from the non-GUP corrected quantum Carnot process. From Equation 42 and Equation 16, we get

*Equation 44:* $\Delta\eta \equiv \frac{\Delta Q}{Q_{AB}}\eta_C = \frac{\frac{\lambda}{2}\left(\frac{1}{\beta_h^2} - \frac{1}{\beta_l^2}\right)}{\frac{1}{2\beta_h}\ln\left(\frac{\gamma_A}{\gamma_B}\right)}\eta_C = \frac{\lambda\beta_h\left(\frac{1}{\beta_h^2} - \frac{1}{\beta_l^2}\right)}{\ln\left(\frac{\gamma_A}{\gamma_B}\right)}\eta_C.$

Hence we get

*Equation 45:* $\frac{\Delta\eta}{\eta_C} = \frac{\lambda\beta_h\left(\frac{1}{\beta_h^2} - \frac{1}{\beta_l^2}\right)}{\ln\left(\frac{\gamma_A}{\gamma_B}\right)} = \lambda k T_h \frac{1 - (\beta_h/\beta_l)^2}{\ln\left(\frac{\gamma_A}{\gamma_B}\right)}$

From our discussion in section 3, we have $\beta_C \gamma_C = \beta_B \gamma_B$. Hence we can write $\frac{\gamma_A}{\gamma_B} = \frac{\beta_h}{\beta_l}\frac{\gamma_A}{\gamma_C}$. Equation 45 becomes

*Equation 46:* $\frac{\Delta\eta}{\eta_C} = \lambda k T_h \frac{(1-r^2)}{\ln(r \cdot r_L)}$ with $r \equiv \frac{\beta_h}{\beta_l} = \frac{T_l}{T_h}$ and $r_L \equiv \frac{L_C^2}{L_A^2}$ with $0 < r < 1$ and $r_L > 1$. Note that we wrote the GUP correction term $\frac{\Delta\eta}{\eta_C}$ in Equation 46 in terms of the highest temperature and width ($T_h$ and $L_C$) and the lowest temperature and width ($T_l$ and $L_A$). From Equation 46, we plot $f \equiv \frac{\Delta\eta}{\eta_C}\left(\frac{1}{\lambda k T_h}\right)$ in Figure 3 and Figure 4. We only consider the parts in both figures where $f > 0$ since as mentioned above, we expect $\Delta\eta > 0$. In

*Figure 3*, we plot $f(r)$ with $r_L = 2$. As r increases $T_l \to T_h$, $\Delta\eta \to 0$ while decreasing r, or lower $T_l$, the GUP correction $\Delta\eta$ increases. Since thermal energy decreases, we expect gravitational effects to predominate. In Figure 4, we plot $f(r_L)$ with $r = 0.5$. As $r_L$ increases (maximum width $L_C$ increases), $\Delta\eta \to 0$ while as $r_L$ decreases (maximum width $L_C$ decreases), $\Delta\eta$ increases. This is expected since the effect of the minimal length is more pronounced for decreasing length dimensions. Choosing different values of $r_L$ to plot and $f(r)$ and different values of $r$ to plot $f(r_L)$ give the same behavior as in *Figure 3* and Figure 4 respectively.

## 7. GUP-corrected Quantum Otto Cycle

Recall the process in an Otto engine: $A \to B$, an isochoric (constant volume) process, $B \to C$, an adiabatic expansion, $C \to D$, an isochoric (constant volume) process and $D \to A$, an adiabatic compression. As before we use Equation 38 to write down the heat involved. For the isochoric processes $A \to B$, and $C \to D$,

*Equation 47:* $Q_{AB}^G = Q_{AB} - \frac{\lambda}{2}\left(\frac{1}{\beta_B^2} - \frac{1}{\beta_A^2}\right)$ and $Q_{CD}^G = Q_{CD} - \frac{\lambda}{2}\left(\frac{1}{\beta_D^2} - \frac{1}{\beta_C^2}\right)$

respectively. For the corresponding "adiabatic" processes $B \to C$ and $D \to A$ yield

*Equation 48:* $Q_{BC}^G = -\frac{\lambda}{2}\left(\frac{1}{\beta_C^2} - \frac{1}{\beta_B^2}\right) = \frac{\lambda}{2}\left(\frac{1}{\beta_B^2} - \frac{1}{\beta_C^2}\right)$ and $Q_{DA}^G = -\frac{\lambda}{2}\left(\frac{1}{\beta_A^2} - \frac{1}{\beta_D^2}\right)$.

With the GUP correction the processes $B \to C$ and $D \to A$ have some heat involved. Since $Q_{AB} > 0$ and $Q_{CD} < 0$, we expect $Q_{AB}^G > 0$ and $Q_{CD}^G < 0$ since the GUP-correction is $\frac{\lambda}{2}\left(\frac{1}{\beta_f^2} - \frac{1}{\beta_i^2}\right)$ is smaller than $Q_{if}$. $Q_{BC}^G > 0$ and $Q_{DA}^G < 0$ since $T_B > T_C$ and $T_A > T_D$ respectively as noted in Figure 2. Hence from Equation 47 and Equation 48

*Equation 49:* $Q_{in}^G = Q_{AB}^G + Q_{BC}^G = Q_{AB} - \frac{\lambda}{2}\left(\frac{1}{\beta_B^2} - \frac{1}{\beta_A^2}\right) + \frac{\lambda}{2}\left(\frac{1}{\beta_B^2} - \frac{1}{\beta_C^2}\right) = Q_{AB} + \frac{\lambda}{2}\left(\frac{1}{\beta_A^2} - \frac{1}{\beta_C^2}\right)$

and

Equation 50: $Q_{out}^G = Q_{CD}^G + Q_{DA}^G = Q_{CD} - \frac{\lambda}{2}\left(\frac{1}{\beta_D^2} - \frac{1}{\beta_C^2}\right) - \frac{\lambda}{2}\left(\frac{1}{\beta_A^2} - \frac{1}{\beta_D^2}\right) = Q_{CD} + \frac{\lambda}{2}\left(\frac{1}{\beta_C^2} - \frac{1}{\beta_A^2}\right) =$
$-|Q_{CD}| + \frac{\lambda}{2}\left(\frac{1}{\beta_C^2} - \frac{1}{\beta_A^2}\right)$.

Putting Equation 49 and Equation 50 into Equation 1, we get $\eta_O^G = \frac{Q_{AB} + \frac{\lambda}{2}\left(\frac{1}{\beta_A^2} - \frac{1}{\beta_C^2}\right) - \left(|Q_{CD}| - \frac{\lambda}{2}\left(\frac{1}{\beta_C^2} - \frac{1}{\beta_A^2}\right)\right)}{Q_{in}^G} =$

$\frac{W}{Q_{in}^G}$ with $W$ as the non-GUP work involved. Manipulating $\eta_O^G$ with Equation 49, we get $\eta_O^G =$

$\frac{W}{Q_{AB} + \Delta Q_{AB}} = \frac{W}{Q_{AB}} \frac{1}{1 + \frac{\Delta Q_{AB}}{Q_{AB}}} \approx \frac{W}{Q_{AB}}\left(1 - \frac{\Delta Q_{AB}}{Q_{AB}}\right)$ where

Equation 51: $\Delta Q_{AB} \equiv \frac{\lambda}{2}\left(\frac{1}{\beta_A^2} - \frac{1}{\beta_C^2}\right)$.

Finally, $\eta_O^G = \eta_O - \frac{W \Delta Q_{AB}}{Q_{AB}^2}$ or

Equation 52: $\eta_O^G = \eta_O - \Delta\eta_O$, $\Delta\eta_O \equiv \frac{W \Delta Q_{AB}}{Q_{AB}^2}$

with $\Delta\eta_O$ as the GUP correction. It is clear that $\Delta\eta_O > 0$ since $W$ and $\Delta Q_{AB}$ are both positive. Hence the GUP-corrected quantum Otto efficiency is less than the non-GUP quantum Otto efficiency.

Similar to the GUP-corrected Carnot cycle let us look at $\frac{\Delta\eta_O}{\eta_O}$. From Equation 25 and Equation 26,

Equation 53: $W = Q_{AB} - |Q_{CD}| = \left(\frac{\gamma_h - \gamma_l}{2}\right)\left(\frac{1}{\beta_h \gamma_h} - \frac{1}{\beta_l \gamma_l}\right)$.

Putting Equation 53, Equation 51, Equation 25 into $\Delta\eta_O$ of Equation 52, and with Equation 30, we get

Equation 54: $\frac{\Delta\eta_O}{\eta_O} = \lambda \beta_h \left(\frac{1}{\beta_A^2} - \frac{1}{\beta_C^2}\right) \frac{1}{1 - r_L^O r}$ where $r_L^O \equiv \frac{\gamma_h}{\gamma_l} = \frac{L_l^2}{L_h^2}$ and $r \equiv \frac{\beta_h}{\beta_l} = \frac{T_l}{T_h}$

with $r_L^O > 1$ and $0 < r < 1$. Since processes BC and DA are not adiabatic when we put in the GUP correction (see Equation 48), Equation 14 does not apply and we cannot relate $\beta_A$ to $\beta_D = \beta_l$ and $\beta_C$ to $\beta_B = \beta_h$. To facilitate an analysis like section 6 above, let us assume that similar to the relations in Equation 24 and Equation 27, we can write generally

Equation 55: $\beta_A = f_{AD}\beta_D = f_{AD}\beta_l$ and $\beta_C = f_{CB}\beta_B = f_{CB}\beta_h$

With $f_{AD} < 1$ and $f_{CB} > 1$ since $T_D < T_A$ and $T_B > T_C$ as noted in Figure 2. For the non-GUP case in Equation 24 and Equation 27, $f_{AD} = \frac{\gamma_l}{\gamma_h}$ since $\gamma_A = \gamma_B = \gamma_h$, $f_{CB} = \frac{\gamma_h}{\gamma_l}$ since $\gamma_C = \gamma_D = \gamma_l$ for the isochoric processes AB and CD. Putting Equation 55 into Equation 54, we get

Equation 56: $\frac{\Delta\eta_O}{\eta_O} = \lambda \frac{\beta_h}{\beta_A^2}\left(1 - \frac{f_{AD}^2}{f_{CB}^2} \cdot \frac{1}{r}\right)\left(\frac{1}{1 - r_L^O r}\right)$.

In Figure 5, we plot Equation 56 $f_{GUP}^O(r) \equiv \frac{\Delta \eta_O}{\eta_O} \cdot \frac{\beta_A^2}{\lambda \beta_h}$, with $f_{AD} = 0.5, f_{CB} = 2$, and $r_L^O = 5$. We only consider the region where $f_{GUP}^O(r) > 0$ since as noted above $\Delta \eta_O > 0$. We consider the region $0.2 < r < 0.25$. The lower the value of $r \equiv \frac{T_l}{T_h}$, (lower value of $T_l$) the higher the value of $\frac{\Delta \eta_O}{\eta_O}$. In Figure 6 we plot $f_{GUP}^O(r_L^O)$ with $f_{AD} = 0.5, f_{CB} = 2$, and $r = 0.1$. Just as before we only consider the region where $f_{GUP}^O(r_L^O) > 0$ since as noted above $\Delta \eta_O > 0$. We consider the region $r_L^O > 10$. As $r_L^O$ increases (maximum width $L_C$ increases), $\Delta \eta_O \to 0$ while as $r_L^O$ decreases (maximum width $L_C$ decreases), $\Delta \eta_O$ increases. Both cases exhibit a similar behavior as the GUP-corrected Carnot engine. One can show generally that these behavior in Figure 5 and Figure 6 occur when $\frac{1}{r_L^O} < r < \frac{f_{AD}}{f_{CB}}$.

## 8. Conclusions

We analyzed the two most commonly studied heat engines in undergraduate physics namely the Carnot and Otto engines. Following the approach of [8] and [9], we calculated the heats involved in the quantum Carnot and Otto cycles and the corresponding GUP-corrected quantum cycles using the partition function of the working substance namely a particle in an infinite square well. Similar to the results of [24], the GUP correction to the entropy (and consequently to the heat energy) is independent of the dimension of the container (in our case the dimension of the well). In both cases the work = heat input minus |heat output| do not change with the GUP correction since the GUP-corrected heat is independent of the width of the well (see Equation 38). The adiabatic processes in both the Carnot and Otto cycles cease to be adiabatic in the GUP-corrected cases. This behavior is expected since the introduction of the GUP correction term in the Hamiltonian of the Schrodinger equation alters the energy of the working substance. A similar behavior is observed in [9] where an additional interaction term was introduced in the Hamiltonian. The GUP correction lowers the efficiency of the quantum heat engines discussed. In this paper (due to a GUP correction) and in [9], (due to an interaction term) we have an additive correction term in the Hamiltonian of the Schrodinger equation. In both cases the correction term resulted in a decrease in the efficiency. For both the Carnot and Otto engines, the GUP effects increase as the temperature of the cold heat bath decreases and as the width of the potential well decreases.

With the general GUP-corrected heat energy calculated in Equation 38 with a particle in an infinite square well as the working substance, we can explore the GUP-correction in other heat engines such as the Szilard engine. It will also be interesting to study the GUP-correction using different working substances such as a two-level system and a harmonic oscillator.

# 9. Figures

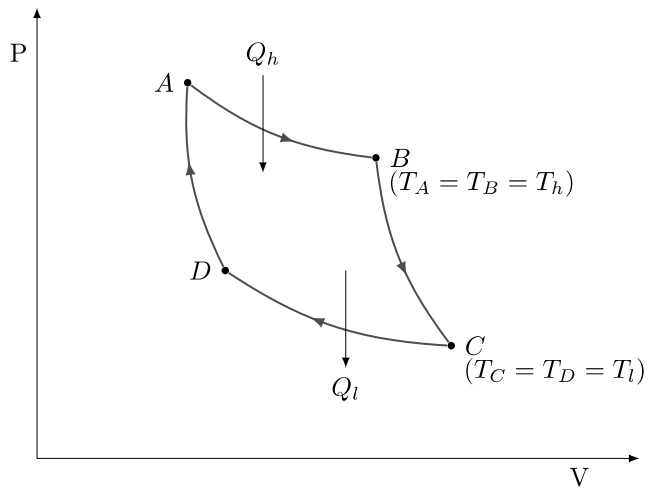

*Figure 1*: Classical Carnot Cycle

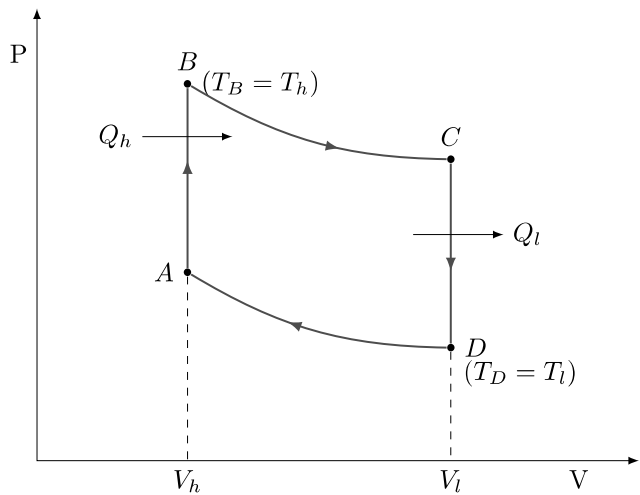

*Figure 2: Classical Otto Cycle with $T_B > T_A > T_C > T_D$*

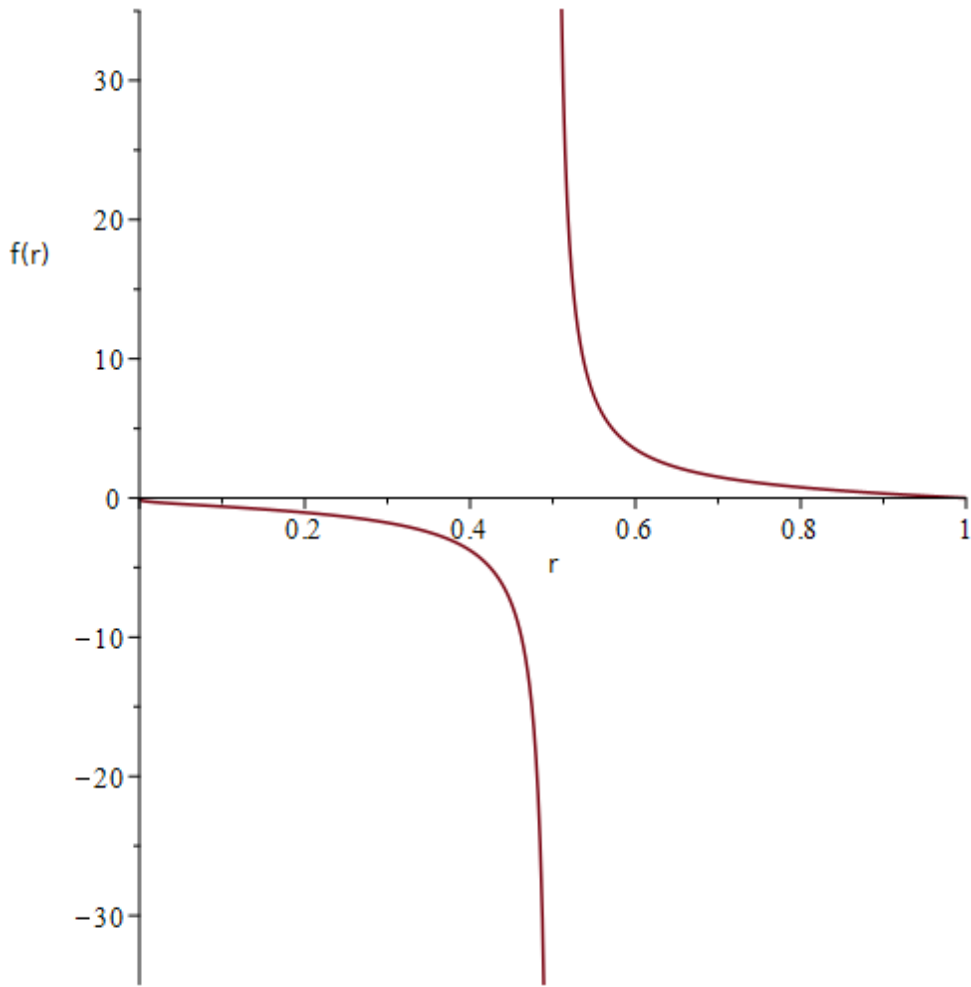

*Figure 3* Plot of $f(r)$ with $r_L = 2$.

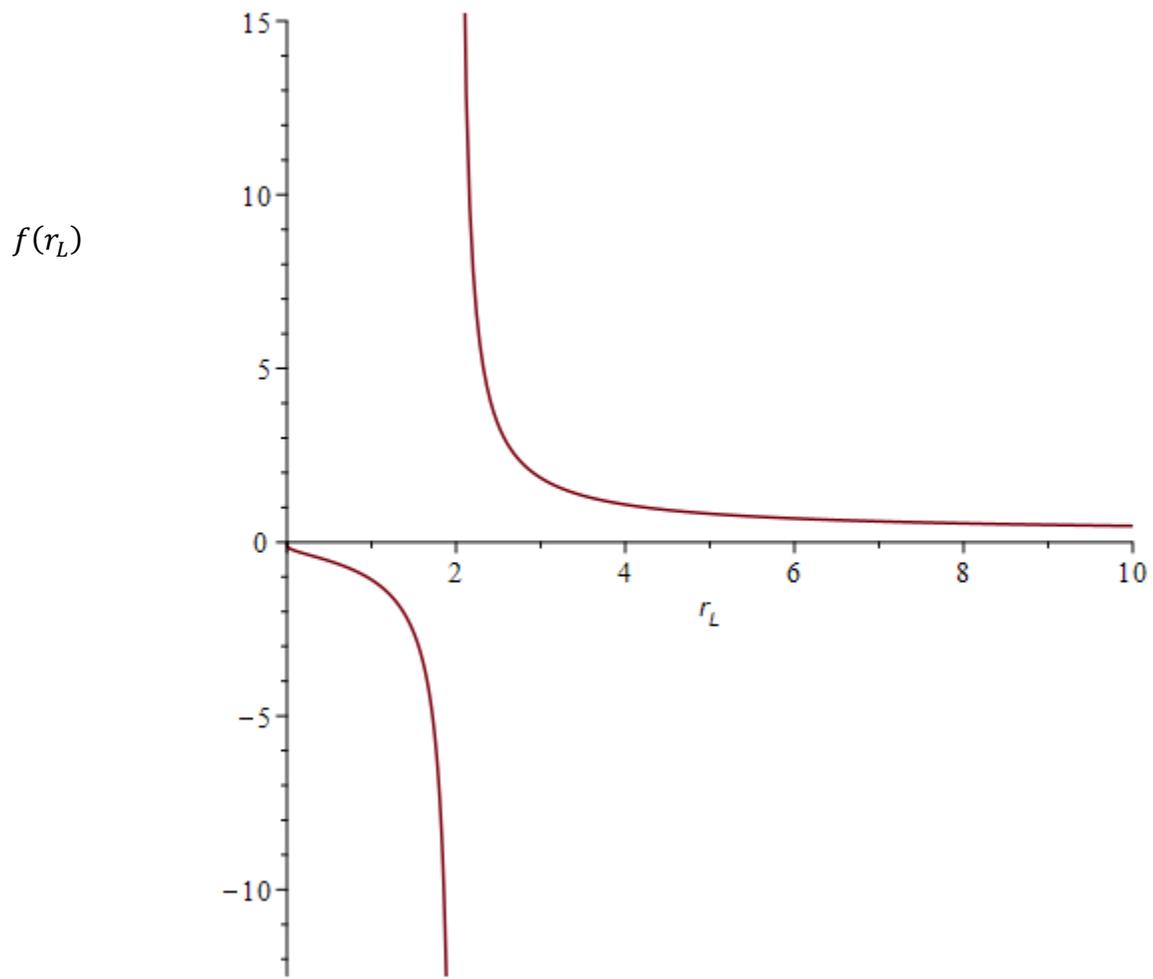

*Figure 4* Plot of $f(r_L)$ with $r = 0.5$.

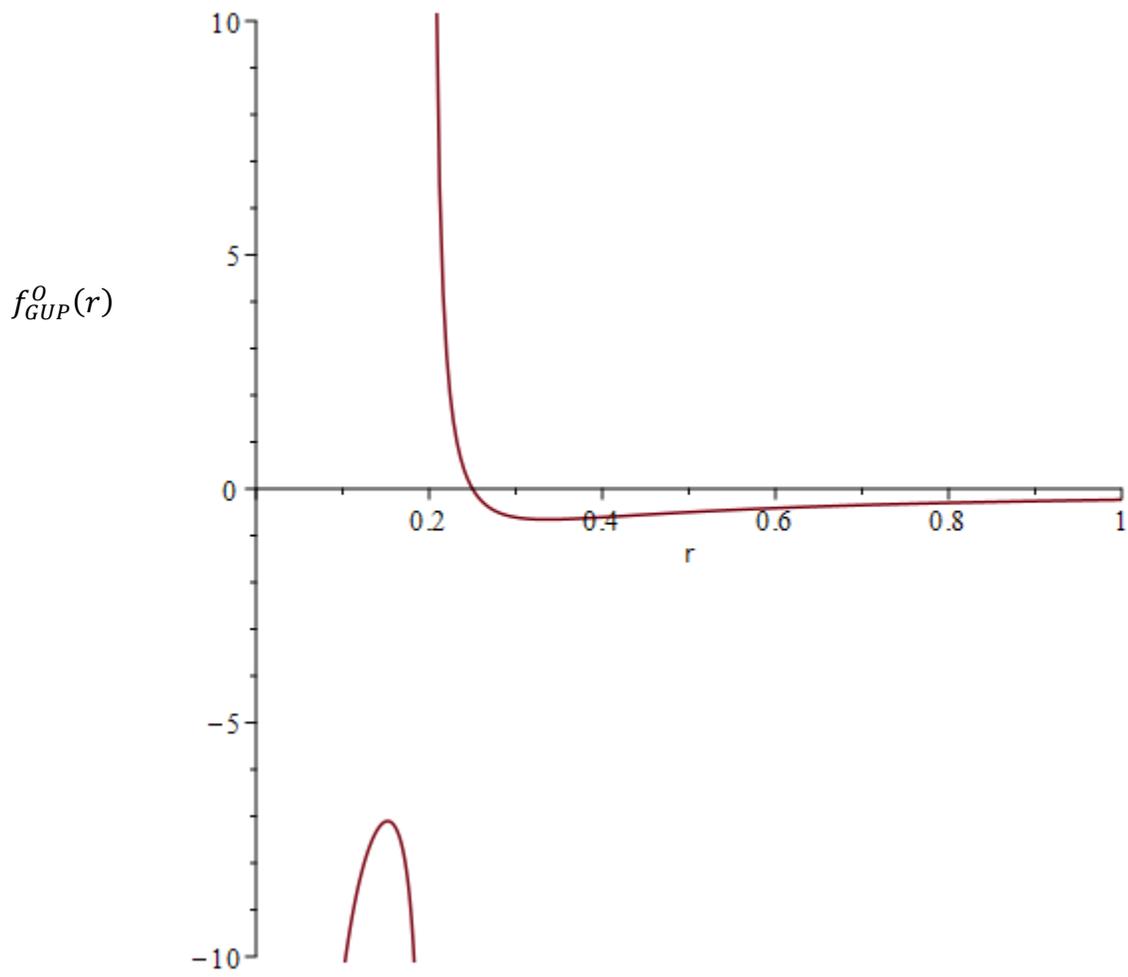

*Figure 5* Plot of $f_{GUP}^O(r)$, with $f_{AD} = 0.5, f_{CB} = 2$, and $r_L^O = 5$.

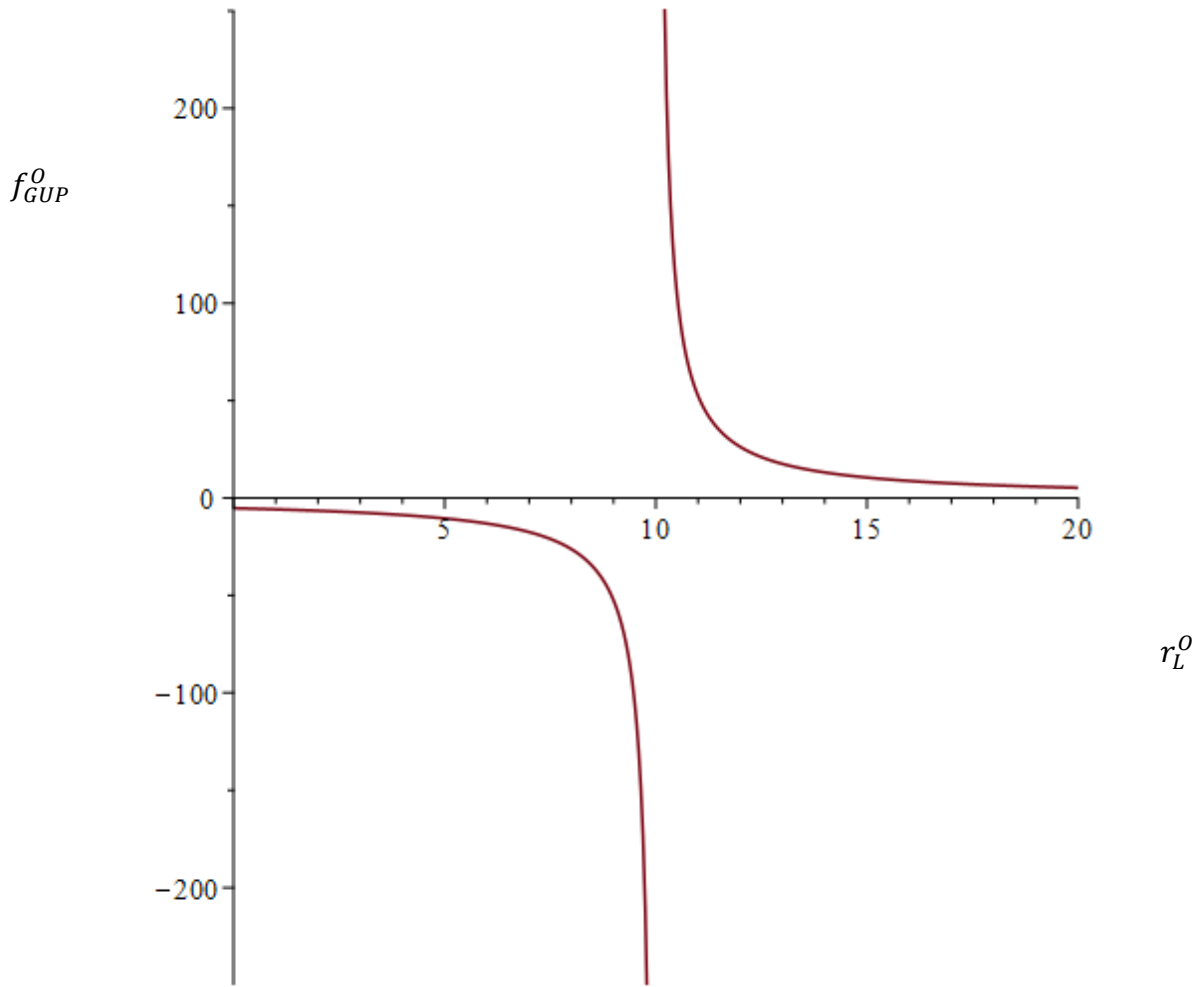

*Figure 6* Plot of $f^O_{GUP}(r^O_L)$ with $f_{AD} = 0.5, f_{CB} = 2$, and $r = 0.1$

**Acknowledgements:**

The authors would like to thank Jonathan Marcel and Sydney Carr for their participation at the beginning of this research project.